\documentclass[a4paper, 10pt]{article}
\usepackage{fullpage}

\long\def\comment#1{}

\newlength{\figtblfootnotemargin}
\newlength{\figtblfootnotewidth}

\newcommand{\SquareOfSize}[2]{
 \fbox{\hsize #1cm \hbox to #1cm{\vbox{#2}}}
}

\newsavebox{\wholeWidthLine}
\sbox{\wholeWidthLine} {\rule[0.1in]{\textwidth}{.01in}}

\newcommand{\bq}{\begin{quote}}
\newcommand{\eq}{\end{quote}}
\newcommand{\be}{\begin{enumerate}}
\newcommand{\ee}{\end{enumerate}}
\newcommand{\bi}{\begin{itemize}}
\newcommand{\ei}{\end{itemize}}
\newcommand{\bie}{\begin{itemize}\begin{enumerate}}
\newcommand{\eie}{\end{enumerate}\end{itemize}}
\newcommand{\ba}{\begin{array}}
\newcommand{\ea}{\end{array}}
\newcommand{\btbl}{\begin{tabular}}
\newcommand{\etbl}{\end{tabular}}
\newcommand{\bequ}{\begin{displaymath}}
\newcommand{\eequ}{\end{displaymath}}
\newcommand{\bequa}{\begin{eqnarray*}}
\newcommand{\eequa}{\end{eqnarray*}}
\newcommand{\bc}{\begin{center}}
\newcommand{\ec}{\end{center}}
\newcommand{\btab}{\begin{tabbing}}
\newcommand{\etab}{\end{tabbing}}

\newcommand{\godown}{ \vspace*{0.3cm}}

\def\mpr#1{\ifmmode #1 \else #1 \fi}

\newcommand{\DPATTRNAME}{{\large{{\bf SC}$_{attr}$}}}
\newcommand{\DPISANAME}{{\large{{\bf SC}$_{isA}$}}}
\newcommand{\DPINNAME}{{\large{{\bf SC}$_{in}$}}}

\usepackage{amssymb,amsfonts,amsmath}
\usepackage{epsfig}
\usepackage{subfig}
\usepackage{url}
\usepackage{algorithm}
\usepackage[noend]{algorithmic}
\usepackage{dsfont}
\usepackage[table]{xcolor}
\usepackage{tabularx}
\usepackage{paralist} 

\urldef{\mails}\path|tzitzik@ics.forth.gr|

\newcommand{\ProbMusic}[0]{{\tt  ProbMusic}}

\begin{document}

\newcommand{\internal}[1]{
  \comment{
      \footnote{
          \fbox{{\tiny InternalNote:}}
          {\tt\em
                #1
          }
      }
  }
}

\title{
A Simple Method to Produce Algorithmic MIDI Music
based on Randomness, Simple Probabilities and Multi-Threading
}


 \author{
    Yannis Tzitzikas \\
    Computer Science Department, University of Crete, Greece, and\\
    Institute of Computer Science, FORTH-ICS, Greece\\
    \mails
    }

\comment{====LNCS====
 \author{
    Yannis Tzitzikas
    }
    \institute{
    Computer Science Department, University of Crete, Greece, and\\
    Institute of Computer Science, FORTH-ICS, Greece\\
    \mails
    }
====}

\maketitle

\begin{abstract}
This paper introduces  a simple method
for producing multichannel MIDI music
that is based on randomness and simple probabilities.
One distinctive feature of the method is that
it produces and sends in parallel to the sound card
more than one unsynchronized channels
by exploiting the multi-threading capabilities of general purpose programming languages.
As consequence the derived sound offers a quite ``full" and ``unpredictable"
acoustic experience to the listener.
Subsequently the paper  reports the results of an evaluation with users.
The results were very surprising:
the majority of users responded that they could tolerate  this music in various occasions.
\end{abstract}

\section{Introduction}

The work presented in this paper
started while we were
investigating a possible programmatic assignment for
the needs of a university course on object-oriented programming,
that involved the usage of a library for MIDI music.
We ended up posing to  ourselves questions like:
\begin{compactitem}
\item  Can we produce algorithmically  music
       which can  be acceptable (or even pleasing) by humans?
       Does this requires having  knowledge background in music?
\item Can we produce potentially  infinite (in length) music
      that is not periodic (and thus not boring)?
\item Can we make humans perceive that such algorithmically-produced music is human-made?
\end{compactitem}
A positive answer to the above questions would be interesting from various perspectives.
A very concrete and practical benefit
is that we would be able  to  produce infinite series of music
without having to care about the space required for storing that music,
since only the algorithm that produces the music would have to be stored.
Another concrete benefit
is that if the composition method
does not require background in music,
then any
user
could easily
compose, or adapt, such mucic to his/her preferences or situation.

\godown


In this paper we present an algorithmic
method for  producing MIDI music semi-randomly.
The composer provides
as input a very short specification
of the desired {\em notes} and their {\em frequency},
the desired {\em instruments},
and the desired {\em octaves and durations}.
An algorithm reads this specification
and by randomly picking elements from that specification,
produces
sequences of MIDI music of the desired length.
One distinctive feature
is that more than one channels are produced
and are sent in parallel to the sound  card,
and this feature  makes the derived mixed music
more interesting and less expected.

Another characteristic of the proposed algorithm
is that it is ``context-free",
i.e. for producing the next score
the algorithm does not consider what it has produced before.
This means that  the only ``memory", or ``pattern", that is used,
is actually the memory of the employed random generator.

\comment{
    The general Research questions
    \begin{compactitem}
    \item Can we produce in this way music that can be heard ?
    \item Is it more pleasing than other kinds of composition?
    \end{compactitem}
==}

The evaluation with users was quite surprising:
almost all of them stated that they could hear such  music
and a significant percentage of them
stated they do like some aspects of that  music.

The rest of this paper is organized as follows:
Section \ref{sec:CompositionProcess} describes the composition process,
i.e. the  specification of the composition,
the semantics of these specifications,
and the algorithm used for producing the sound.
Section \ref{sec:Implementation} describes the
system  \ProbMusic\ that realizes the proposed method.
Section \ref{sec:Evaluation} reports the results of an evaluation with users.
Finally Section \ref{sec:Conclusion} concludes the paper.

\section{The Composition Process}
\label{sec:CompositionProcess}

The process consists of two steps.
In the first one, the composer provides a small specification
consisting of the desired notes/octaves/instruments
(this is analyzed in \S \ref{sec:CompoSpec}).

In the second, an algorithm reads this specification,
picks randomly  elements (notes/octaves/instruments)
and generates one MIDI sequence.
Let call this  sequence {\em mscore}.
Specifically the algorithm produces
{\em three}
mscores, say $mscore_1, mscore_2$ and $mscore_3$.
These three scores are not identical,
because each one was produced by the process described before,
therefore it can be different due to the random selection,
which is based on a generator of random numbers.

After having produced these three mscores,
{\em three threads}  are  produced,
by exploiting the multi-threading capabilities\footnote{
    http://en.wikipedia.org/wiki/Thread\_(computing)
} of the adopted
programming language (in our case we use the Java \cite{gosling2000java}\footnote{
    http://en.wikipedia.org/wiki/Java\_(programming\_language)
} programming language).
Each thread is responsible for playing one {\em mscore},
however the three threads do not start at the same time:
the second threads starts 3 seconds after the beginning of the first thread,
while the third thread  starts 3 seconds after the beginning of the second.
It follows that after around 6 seconds,
the produced sound
is the result of mixing the sounds of three parallel MIDI channels.
It is also worth noting that the threads are not enforced to be synchronized.
This means that the hosting computer system
(virtual machine, operating system, load of the machine at that time)
affects the priorities of the threads
and consequently the sound that will be reproduced at each point in time.

It follows that the produced  sound has {\em three aspects of randomness or non-determinism}: \\
a) the random selection of elements from what the composer has specified, \\
b) each of the three mscores is different because they are based on different
 sequences of random numbers, and  \\
c) the priority (and schedule of execution) of the threads on the machine of user
is not deterministic. \\
As a consequence, the same composition specification
sounds differently each time  it is requested.

\subsection{Specification of Composition}
\label{sec:CompoSpec}


An example of a specification is given in Figure \ref{fig:SpecExample}.

\begin{figure*}[phtb]
\begin{verbatim}
{
   {"Relaxing,  Oct 24, 2013"},
   {"A","C","E","G"},
   {"3q","2h","5w","h","4h"},
   {"Oboe","ELECTRIC_JAZZ_GUITAR","Atmosphere","Choir","Choir_AAHS"},
}
\end{verbatim}
\caption{Example of the specification of a musical piece}
\label{fig:SpecExample}
\end{figure*}

\comment{========
    \begin{verbatim}
    {
       {"Relaxing,  Oct 24, 2013"},
       {"A","C","E","G"},
       {"3q","2h","5w","h","4h"},
       {"Oboe","ELECTRIC_JAZZ_GUITAR","Atmosphere","Choir","Choir_AAHS"},
    },
    \end{verbatim}
===}

The first row is the name of the composition and the date of its composition.
The second row is a {\em bag of notes},
here  [A, C, E, G].
The algorithm will pick randomly notes from this bag.
We call it bag, instead of set,
because duplicates are allowed
meaning that if the composer
provides a row of the form  [A C C],
the algorithm
with 1/3 probability will be picking the note A and with probability 2/3 the note C.

The third row is a bag of
words consisting of one or two letters.
The first letter indicates  {\em octave}, while the second indicates {\em duration}.
As regards duration, the possible values (and their meaning) follow:
\begin{compactitem}
\item  w: whole duration
\item  h: half duration
\item  q: quarter duration
\item  i: eighth duration
\end{compactitem}
The digits denote octaves and range from 1 to 10.
\comment{
    \begin{compactitem}
    \item  1:
    \item  2:
    \item  3:
    \item  4:
    \item  5:
    \end{compactitem}
}

The fourth row is a {\em bag of instrument names}, the instruments
that the composer prefers for this composition.

We should mention
that all rows are {\em bags} (not sets)
%
meaning that the composer can specify the desired probabilities
(of notes, duration-octave pairs, instruments)
through  duplicating accordingly the desired values.

\subsection{Semantics of Composition}
\label{sec:CompoSem}

The algorithm picks
elements
from the bags and forms what we  call  {\em words}.
A word contains a randomly picked note,
and a randomly picked duration-octave element.
As regards instruments,
the algorithm after
the production of a word
changes the instrument
(by picking randomly an instrument
from the provided bag)
with probability 40\%
(i.e. with 60\% probability the instrument does not change).

Figure \ref{fig:ConsoleOutputExample}
shows the console output
of the algorithm.
In this example, we have requested {\em music length} equal to 33.
This means that  each of the three threads,
will derive an $mstring$
consisting of 33 words.
Since we create three threads,
the listener will hear  99 (=3 * 33) words.
The console output also shows the
times when each of the three threads starts.

\begin{figure*}[phtb]
{\small
\begin{verbatim}
Thread No0 has started on 2013/10/28 00:43:12
T[Allegro] I[Choir]  A3q A2h I[Oboe]  Eh A5w I[Atmosphere]  C4h I[Choir_AAHS]
E2h Ah Eh I[Choir_AAHS]  E4h Ah I[ELECTRIC_JAZZ_GUITAR]  A4h I[Atmosphere]
E4h Ch A3q G2h Ch I[Choir]  G3q C4h C4h G5w I[ELECTRIC_JAZZ_GUITAR]  E4h E5w
I[Choir_AAHS]  E5w E5w E5w G3q C3q E5w G5w I[Atmosphere]  Eh G3q A3q Ah

Thread No1 has started on 2013/10/28 00:43:15
T[Allegro] I[Oboe]  C3q Ah I[Oboe]  G5w I[Oboe]  A5w I[Atmosphere]  G2h G5w
Eh A4h I[Choir]  G4h I[ELECTRIC_JAZZ_GUITAR]  C5w E4h I[ELECTRIC_JAZZ_GUITAR]
A4h I[Oboe]  C5w A5w C5w G4h I[Choir_AAHS]  A4h I[Oboe]  G5w C5w C3q Gh C4h
Eh I[Atmosphere]  G3q I[Choir_AAHS]  A2h Ch Ch G5w I[ELECTRIC_JAZZ_GUITAR]
Ah I[Oboe]  C5w E5w E5w C3q

Thread No2 has started on 2013/10/28 00:43:18
T[Allegro] I[Oboe]  Gh E4h I[Atmosphere]  E5w I[ELECTRIC_JAZZ_GUITAR]  A5w
 I[Choir]  G5w A3q E4h I[Choir]  C5w I[Atmosphere]  A5w I[Choir]  Ch
 I[Choir] E4h I[Choir]  G5w E3q A4h I[Atmosphere]  C4h A3q G2h A2h I[Choir_AAHS]
 A2h Eh G2h C3q E5w I[ELECTRIC_JAZZ_GUITAR]  A4h Ah E5w C4h E5w
 I[ELECTRIC_JAZZ_GUITAR]  Gh A2h A4h G3q I[Atmosphere]  C5w
\end{verbatim}
}
\caption{Example of the console output}
\label{fig:ConsoleOutputExample}
\end{figure*}

\godown

Note that the configuration and the algorithm that uses it,
is generic in the sense that it can accommodate
various options.
Specifically,
instead of single notes one can have sequences of notes,
e.g. \{"A B","A C","E","G"\}.

\internal{We could have bigger words in the config.
We have selected the simplest possible.
To see if with the fundamental musical elements
one (even user with no background in music)
can come up with something that is acceptable.
}

\subsection{The Role of the Composer}

The composer's role is to produce a specification
like that of Figure \ref{fig:SpecExample}.
For deriving the specification
of the pieces in \cite{yannisYouTubeDec092013},
the author has followed a
straightforward trial-and-error methodology.
This requires selecting a few notes,
some octave-duration pairs, some instruments,
and then test by hearing the result.
In average  each composition
(at least those in \cite{yannisYouTubeDec092013})
required a trial-and-error process that lasted about 20 minutes.

\subsection{The Multiplicity of Distinct Serializations}

The number of different sounds that can be produced
by such a small specifications, like that of Figure \ref{fig:SpecExample},
is very big.
Let $N$ denote the set of distinct notes of a specification (2nd row of a specification),
$OD$ denote the set of distinct octave-duration pairs (3rd row),
and
$I$ the set of distinct instruments (4th row).
It follows that the set of different {\em words} $W$
that can be generated
are
$|N|*|OD|*|I|$ in size.
For the specification of Figure \ref{fig:SpecExample},
we have
$|N|=4$,
$|OD|=5$,
$|I|=5$,
so we have $|W|=100$ (= 4 * 5 * 5) different words
(i.e. $|W| = |N|*|OD|*|I|)$.

The size (as number of words) of each {\em mscore}
is provided by the user.
A default value (that yields music lasting for about 2 minutes and some seconds)
is 120.
Let use $MS$ to denote the number of  words in a mscore.
Since each word is picked randomly,
the number of different mscores of size $MS$
is $|W|^{MS}$.
In our example it is $100^{120}$.
Since the music is produced by three mscores which are played concurrently,
the number of all different triples
of mscores
is $100^{120} \times  100^{120}  \times 100^{120} = 100^{360}$.
Obviously, this number is extremely big
(note that  the number of atoms
in the entire observable universe
is estimated\footnote{
    http://en.wikipedia.org/wiki/Observable\_universe  (Dec 11, 2013)
}
to be around $10^{80}$).
Overall, every time a user
requests to hear the piece of Figure \ref{fig:SpecExample},
he gets back  one of the $100^{360}$ possible serializations.

\comment{== PROBABILITY OF HEARING ONE PIECE MORE THAN ONCE
    This makes each ``reproduction" quite unique:
    consider a user that loves that song
    and for this reason he hears it 10 times per day
    for the next 30 years
    (i.e. 10*365*30=  109,500 times in total).
    The probability of  hearing exactly the same
    sound twice is ...
====}

\subsection{Placement of this Method  in the Landscape}

Let describe the ``coordinates" of the described method of music production
in the landscape of algorithmic music.
\begin{compactitem}
\item As regards the outcome, this method produces polyphonic music,
due to multi-threading.
\item
It is not music composed by the computer, since the composition
specification is given by the user
(in future we would like to investigate deriving the
composition specification without human intervention).
\item
As regards the compositional process,
the algorithm  provides two levels of notational information:
the composition specification
and the three  series as shown in Figure \ref{fig:ConsoleOutputExample}
(however the latter is only one possible ``interpretation"
of the composition specification).
\item
As regards the structure of the compositional algorithm per se,
it is based on a simple mathematical model
(it does not use any  knowledge,  grammar, evolutionary method,
or learning).
\end{compactitem}


\section{Implementation (\ProbMusic)}
\label{sec:Implementation}

The described method and algorithm
for producing multi-threaded music
has been implemented  in a system
that is called \ProbMusic.
It is written using the Java programming language,
while for controlling the MIDI player
it uses the  {\tt jfugue} library
\cite{koelle2007jfugue}.

\begin{figure*}[phtb]
  \centerline{
    \epsfig{figure=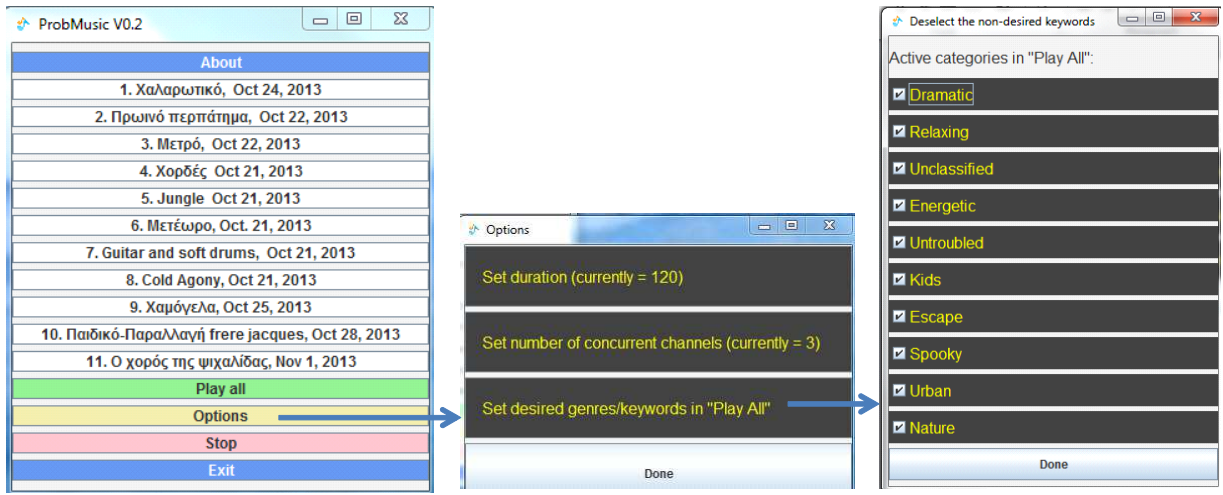,width=16cm}
   }
\caption{The graphical user interface of \ProbMusic}
\label{fig:gui}
\end{figure*}

\ProbMusic\
also offers a kind of ``playlist player" functionality
allowing the user
the select the desired piece (also offering a  ``play all" feature).
Figure \ref{fig:gui} shows a screenshot from its current version
offering 11 musical pieces.
The player provides options for specifying the desired duration
of each piece (expressed as  number of words),
and the number of threads (the default is 3).

Each composition can be annotated (currently only by the composer)
by one or more keywords.
The user can then  request the exclusion  from the ``play all" functionality
of the pieces that contain
non-preferred (for that moment) keywords by unchecking
the corresponding checkbox(es).
Finally, and  for obtaining smooth transition in the ``play all" functionality,
the volume of the sound is gradually decreased at the end of each musical piece.

The entire software is packaged as a single .jar file
that can run in any machine that has the Java Runtime Environment.
The size of the package is  very small,
around 1.1 MB,
comparable to the size of a single mp3 file!

A YouTube video
the contains around 20 seconds
of 9 musical pieces is accessible from
\cite{yannisYouTubeDec092013}.

\section{Evaluation with Users}
\label{sec:Evaluation}

There are numerous works that propose methods for
producing music algorithmically
dating back to Pythagoras
(for various links or categorizations
see \cite{langston1989six,moroni2000vox,wooller2005framework,bell2011algorithmic}),
including attempts  for producing music using very simple methods,
e.g. \cite{heikkila2011discovering}.
However, to the best of our knowledge,
 most works do not report any kind
of evaluation from users,
so it is hard to understand
whether the users liked or not the music produced by such methods.
In our case,
we decided to conduct an evaluation with users  in order to
check whether
the produced music is
acceptable, and if yes to what degree.

In this evaluation,  41 persons participated with ages ranging from 19 to 29.
They were persons either  working at FORTH (computer science researchers, engineers, graduate students)
or undergraduate students of the Computer Science Department of the University of Crete
(specifically  students attending the 2nd year's course ``CS252 Object-Oriented Programming",
Fall 2013-2014).

Each participant was asked
to download the software,
to hear each of the provided pieces (more than once),
and then
to answer a questionnaire.
The software included 6 musical pieces.

The questionnaire contained the following questions:
\bq
{\bf\em\small
\be
\item Select the two pieces that you liked most.
\item 
      Select one of the following choices:
\be
    \item  I cannot tolerate this music.
    \item  Yes, I could hear to this music in some occasions
            (e.g. in public transport means, when I wash dishes, ...).
    \item  Yes, in general I could hear this music
    \item I could say that I like some aspects of this music.
    \item I like this music.
\ee
\item
    Would you be interested in composing music without having any background in music?
\ee
}
\eq

\begin{figure*}[phtb]
  \centerline{
   \fbox{
    \epsfig{figure=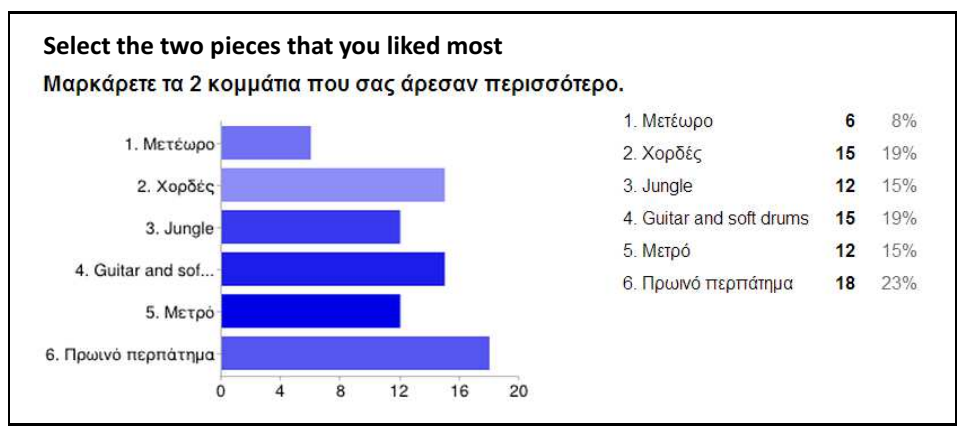,width=12cm}
   }
   }
\caption{Preferences.}
\label{fig:preferred}
\end{figure*}

Figure \ref{fig:preferred} shows the results
regarding the first question
where each participant had to select the two
most preferred pieces.
We do not observe big differences,
since the  percentages range from 8\% to 23\%,
i.e. there was not any particular piece that most participants preferred a lot,
or disliked a lot.

\begin{figure*}[phtb]
  \centerline{
   \fbox{
    \epsfig{figure=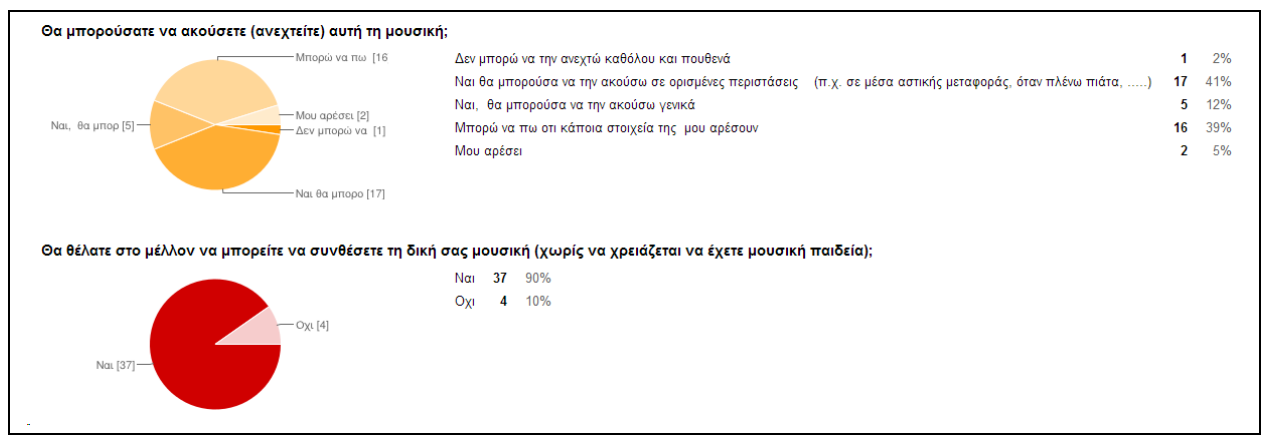,width=16cm}
   }
   }
\caption{The results of the questionnaire.}
\label{fig:survey}
\end{figure*}

Figure \ref{fig:survey} shows the results
of the second the third questions.
As regards the second question, we can see that:
\begin{compactitem}
\item
41\% of the participants said that they  could hear such music in specific cases,
\item
39\% that they liked some aspects of that mucic,
\item
12\% that could hear it in general, and
\item
5\% stated that they like it.
\end{compactitem}
In total,  98\% of the participants declared that they can  tolerate this music
(only 2\% stated that they cannot tolerate  it).
The percentage of persons
that expressed something positive
(liked some aspects, could hear it in general, or even like it)
is 56\%  (=5+12+39).
It was rather surprising to see such a high percentage of positive reactions.

As regards the last question,
90\%  of the participants responded
that they  would like the ability
to produce their own music.

\comment{== PREVIOUS
        In this evaluation, {\bf 17} persons
        participated with ages ranging from 19 to 29.

        Figure \ref{fig:survey} shows the results.
        As regards the second question,
        we can see that
        53\% of the participants said that they  could hear that music in specific cases,
        24\% that they liked some aspects,
        and
        2\% that could hear it in general.

        In total
        it seems that 79\%
        of the participants
        could tolerate it in particular cases.
==============}


\section{Synopsis}
\label{sec:Conclusion}

This paper  presented a  simple
method  for producing multichannel MIDI music.
The method is based on random selection,
very simple probabilities,
and exploits the multi-threading capabilities of programming languages.
It does not exploit any result  of the music theory.
It is surprising
that the produced music
was so well accepted by the users.
Almost all of them (98\%) stated that they could hear such  music
and a significant percentage of them
stated they do like some aspects of that  music.

Since the composition is based on a very simple input,
simple users (i.e. users  with no background in music),
could in future use this method for composing their own music.
In future we would like to study which configuration
``patterns" lead to aesthetically pleasing music,
for
attempting to provide
an automatic method for producing
even composition specifications.
It would also be interesting
to  compare  with users
the produced music
with the music produced by other algorithmic methods.

The system \ProbMusic\  is free
and is available upon request.

\internal{
    BTW:
    is there any other approach (ideally popular) that can produce
    infinite in length music?
    Which are the competitors?
}

\subsection*{Acknowledgements}

Many thanks to  Panagiotis Papadakos for informing  me about the existence
of the {\tt jfugue} library,
and to the students of the course ``CS252 Object-Oriented Programming" (Fall 2013-2014)
of the Computer Science Department of the University of Crete,
for participating to the evaluation.

%

\bibliographystyle{abbrv}
\bibliography{music}

\end{document}